# Single Cell Transcriptome Research in Human Placenta

Hui Li[1], Qianhui Huang[2], Yu Liu[1], Lana X Garmire[1, $]

[1]Department of Computational Medicine and Bioinformatics, University of Michigan, Ann Arbor, MI, USA

[2]Department of Biostatistics, School of Public Health, University of Michigan, Ann Arbor, MI, USA

$Corresponding author: lgarmire@med.umich.edu

**Abstract**

Human placenta is a complex and heterogeneous organ interfacing between the mother and the fetus that supports fetal development. Alterations to placental structural components are associated with various pregnancy complications. To reveal the heterogeneity among various placenta cell types in normal and diseased placentas, as well as elucidate molecular interactions within a population of placental cells, a new genomics technology called single cell RNA-Seq (or scRNA-seq) has been employed in the last couple of years. Here we review the principles of scRNA-seq technology, and summarize the recent human placenta studies at scRNA-seq level across gestational ages as well as in pregnancy complications such as preterm birth and preeclampsia. We list the computational analysis platforms and resources available for the public use. Lastly, we discuss the future areas of interest for placenta single cell studies, as well as the data analytics needed to accomplish them.

**Introduction**

The placenta is the interface between the mother and the fetus, which mediates the exchange of gas, nutrients, waste, and produces hormones and growth factors that support fetal development and ensure a healthy pregnancy. Much of the understanding of early stages of human implantation and placental development is based on histological analyses of specimens of the Boyd Collection and Carnegie Institution of Washington (Hertig 1956; Hamilton *et al*.1960), as well as anatomical studies of species closest to humans (Enders 2007). In addition, in vitro fertilization technology (Deglincerti *et al*. 2016; Shahbazi *et al*. 2016) and pre-implantation in the mouse (Cockburn *et al*. 2010) have contributed to our knowledge of pre-implantation events in humans. The development of the human placenta starts from the formation of the trophectoderm and the inner cell mass. At this developmental stage, the pre-implantation embryo is referred to as a blastocyst.

Implantation starts around day 7 post conception (p.c.) when the blastocyst attaches and adheres to the uterine epithelium (Hertig 1956). How this is achieved is not clear. Two groups cultured human embryos *in vitro* for 12–13 days p.c. and unveiled the self-organizing abilities in vitro attached human embryos (Deglincerti *et al*. 2016; Shahbazi *et al*. 2016). The presence of cell adhesion molecules including integrin, E-cadherin and L-selectin, on human oocytes, early embryos, and blastocysts, suggests that these molecules may play a role in embryo attachment and adhesion (Campbell *et al*. 1995). Following blastocyst attachment and adhesion, trophoblast cells undergo cell fusion to form the multinucleated syncytiotrophoblast (SCT), which invades the maternal uterine stroma. The mechanisms underlying the transition of cytotrophoblasts into SCTs remains largely unknown. The blastocyst eventually embeds itself into the stromal vasculature of the uterine lining (Norwitz *et al*. 2001; Boyd *et al*. 1970). With the embryo implanted in the uterus, epiblast and endoderm cells cavitate to form the amniotic cavity and yolk sac, respectively (Enders 1986). Around day 13 p.c., the cytotrophoblast cells underlying the SCT proliferate in columns and penetrate the cord of SCT, forming primary villi. Two days later, a connective tissue core derived from the extraembryonic mesenchyme invades the primary villi, transforming them into secondary villi (Boyd *et al*. 1970). Fetal blood vessels begin to form in the villi core by day 20 p.c., marking the formation of tertiary villi, the first generation of which are the mesenchymal villi. These stages of the development of new villi are repeated throughout pregnancy. From this time onwards, placental villi are tertiary villi consisting of a vascular network, mesenchyme, cytotrophoblasts and SCTs. These 4 constituents together form the placental barrier (Boyd *et al*. 1970). Around the 5th week p.c., mesenchymal villi begin to differentiate into immature intermediate villi with increased villous diameter and appearance of stromal channels, and later into stem villi by means of central stromal fibrosis (Castellucci *et al*. 1990). From around the 23th week p.c. until term, the mesenchymal villi differentiate into mature intermediate villi, from which highly capillarized terminal villi arise. These terminal villi, which begin to appear at around the 25th week p.c. and account for nearly 40% of villous volume of the placenta at term, are the most effective structures for fetal-maternal diffusion exchange (Castellucci et al, 2000). Morphometric observations have shown that, although the villous growth slows down in late pregnancy, it continues to grow towards term (Boyed, 1984). If the maternal environment becomes unfavorable, the villous will continue branching past term.

Structurally, the placenta is a complex and heterogeneous organ consisting of multiple different cell types that carry out varied functions. The functional unit of the placenta is the chorionic villus that consists of a stromal core, an inner layer of villous cytotrophoblasts (VCT) and an outer layer of multinucleated SCTs that cover the surface of the villous tree. The stromal core contains a range of cells including macrophages (also called Hofbauer cells), mesenchymal stromal cells, fibroblasts, and fetal endothelial cells (Figure 1). Hofbauer cells are placental villous macrophages of fetal origin. scRNA-seq of the first-trimester placenta shows there are at least two subtypes of Hofbauer cells (Liu *et al*. 2018), which express genes involved in maintaining host defense, placental morphogenesis and homeostasis (Seval *et al.* 2007; Liu *et al*. 2018). Two populations of mesenchymal stromal cells were also identified from the single cell transcriptional profiling of first-trimester placental cells (Liu *et al*. 2018). One cell population likely plays a role in the regulation of cell adhesion and migration, whereas the other population in the development of blood vessels and the mesenchyme (Liu *et al*. 2018). Three fibroblast subtypes are identified by scRNA-seq analysis from the first-trimester placenta. Genes highly expressed in these cells are involved in endocrine signaling, endothelial cell migration and angiogenesis, and sodium and fluid homeostasis (Suryawanshi *et al*. 2018). Endothelial cells play an important physiological role in maintaining a vasodilated state in the fetoplacental vessels through the release of vasoactive substances, including nitric oxide. The villous cytotrophoblast (VCT) has been considered the proliferative stem-like cells. They either fuse to form SCTs during placental villous formation or penetrate the SCT to form extravillous trophoblast (EVT) columns in anchoring villi. scRNA-seq of the first-trimester placenta shows that one population of VCT cells express high levels of potentially important fusion genes, indicating that they may be the fusion-competent VCT cells (Liu *et al*. 2018). Single cell gene expression analysis of the first- and third trimester VCT cells indicates the continuous process of VCT to EVT differentiation throughout the pregnancy (Tsang *et al*. (2017; Suryawanshi *et al*. 2018). Invasion of the spiral arteries by EVTs leads to placental vascular remodeling in the early stages of the implantation process. The mechanisms involved in this evasion are still largely unknown. Recent scRNA-seq studies have shown that both the first-and second-trimester EVTs expressed genes involved in invasion and immunomodulation, consistent with a likely role for EVTs in cellular invasion and anti-inflammatory in the early stage (Liu *et al*. 2018; Suryawanshi *et al*. 2018; Vento-Tormo *et al*. 2018). The SCTs cover the entire surface of villous trees and line in the intervillous space, where the maternal-fetal exchange of gas and nutrients occurs. scRNA-seq profiling revealed that these cells expressed genes involved in the production of gestational hormones, which support the fetal and placental development (Tsang *et al*. (2017; Suryawanshi *et al*. 2018). Furthermore, the SCTs are involved in the immune tolerance of the fetus by the maternal immune system.

Other cells may also exist at the interfacing of maternal-fetal surface. Uterine natural killer cells (uNK) constitute a major lymphocyte population in the uterus during early pregnancy. Their number increases around the time of implantation and remains high during the first trimester of pregnancy. uNK cells play a role in protecting the conceptus from maternal immune rejection and have been implicated in placental vascular remodeling (Sojka 2019). Decidual cells are a morphologically distinct cell class observed in the endometrium during implantation and early pregnancy. Functions of decidual cells are thought to regulate the trophoblast invasion and provide nutrition to the embryo (Kearns et al. 1983).

The placenta plays a pivotal role in maintaining the health of both the fetus and the mother. Normal growth and development of placental structural components ensures a healthy pregnancy outcome, whereas alterations to placental structural components are associated with various pregnancy complications such as intrauterine growth restriction (IUGR), preeclampsia and preterm birth. These pregnancy complications and syndromes are associated with increased fetal and neonatal morbidity and have long-term negative impacts on the health of newborns. IUGR refers to a condition in which the fetus does not achieve the expected in utero growth potential. Fetal growth is a complex process that involves genetic and environmental factors, nutrient availability from the mother, as well as hormones and growth factors. The complications of IUGR include perinatal asphyxia, hypoglycemia, hypothermia, cerebral palsy, and pulmonary hypertension of the newborn (Mandruzzato *et al.* 2008). IUGR has many possible causes. A common cause is maternal factors such as maternal health, age, infection and behavioral habits. IUGR can also occur as the result of fetal malformations and chromosomal abnormalities. Preeclampsia is a medical syndrome that typically occurs after 20 weeks of gestation in pregnant women with previously normal blood pressure and affects both the mother and the unborn baby. Preeclampsia is characterized by hypertension and the presence of protein (0.3 g or more of protein in a 24-hour urine collection) in the urine in the mother (Steegers *et al*. 2010). While the etiology of preeclampsia is not fully understood, risk factors are proposed including hypertension and kidney disease, diabetes, obesity, a history of preeclampsia, aberrant placental implantation, and environmental and genetic factors. Preeclampsia has been linked to IUGR, as both conditions are assumed to be related to the similar placental disorder due to reduced trophoblast invasion (Kaufman *et al*. 2003). In fact, IUGR has been included among diagnostic criteria for severe preeclampsia (Tranquilli *et al*. 2014). The only effective treatment for preeclampsia is delivery of the fetus and placenta, leaving this disease one of the leading causes of preterm birth. Since preeclampsia is clinically and molecularly heterogeneous, advances in omics technologies, such as transcriptomics, epigenomics, proteomics and metabolomics, may help capture the complexity of the molecular mechanism of preeclampsia (Ching *et al.* 2014, 2015; Benny *et al.* 2020).

Spontaneous preterm birth is defined as birth before 37 weeks of gestation, with an estimated 15 million babies born preterm every year. It is the leading cause of death among children under 5 years of age and is associated with an increased incidence of developmental delay, cerebral palsy, and pulmonary and neurological disorders. The possible causes of preterm birth include maternal infections, decidual hemorrhage, multiple pregnancies, and chronic conditions such as hypertension and diabetes. It is important to know, that phenotypes such as IUGR, preeclampsia and preterm birth share comorbidities. Special efforts should be thus be paid to control for confounding in data analysis, in order to accurately ping-point the genomics-phenotype associations are truly due to the condition of interest, rather than other confounding factors.

Although the etiology and pathogenesis of these above mentioned diseases are still not clear, the placental dysfunction is thought to be one of the main causes (Burton *et al*. 2018). Normal placental function is dependent on appropriate growth and development of its cells, which are heterogeneous, dynamic, and are determined by the precise regulation of gene expression. Consequently, alterations to placental gene expression are thought to be a major contributor to pregnancy pathologies. Previously, studies investigating the variation in bulk gene expression in placental tissues have been conducted in order to elucidate the molecular basis of placental development and pregnancy pathologies (Chappell *et al*. 2006; Sood *et al.* 2006). However, because the placenta is a complex and heterogeneous organ consisting of multiple cell types that have very different gene expression profiles, these bulk RNA analyses may miss transcriptional differences within a population of placental cells, and more importantly, they are unable to decipher the interactions among these cell types. Studying cellular heterogeneity is therefore critical to understand placenta cellular function in normal and disease states. Fortunately, recent advances in single-cell genomics, especially RNA sequencing (scRNA-seq) technologies, have enabled the characterization of gene expression of various cell types in placenta, simultaneously at single-cell resolution.

**Single-cell RNA sequencing technologies**

Since the first scRNA-seq method was published in 2009 (Tang *et al.* 2009), this technology has been widely used to study dynamic gene expression at the single-cell level in diverse human tissues. The general workflow of scRNA-seq includes single-cell isolation, scRNA-seq library preparation, sequencing and computational analysis (Figure 1). Below we summarize and discuss several popular scRNA-seq methods.

The first, and perhaps most important step in scRNA-seq is the isolation of viable, individual cells from fresh tissues. This can be technically challenging for complex tissues like the placenta, which consists of multiple cell types with various morphological and chemical properties. Methods for isolating cells include micromanipulation, limiting dilution, laser capture microdissection (LCM), fluorescence-activated cell sorting (FACS), magnetic associated cell separation (MACS), and microfluidics, each with its own strengths and weaknesses. Micromanipulation technique uses microscope-guided micropipettes to pick cells from samples with a small number of cells, such as early embryos (Guo *et al.* 2017). Micromanipulation is a simple and robust method, however, its throughput is low as cell transfer from the medium to tubes or plates increases the amount of time. Limiting dilution uses hand-pipettes to isolate single cells by the dilution of cell suspension (Fuller *et al.* 2001). This process is simple and cost-effective, but it is not as efficient as the probability of obtaining a single-cell based on the Poisson distribution. LCM utilizes a laser system aided by a computer to isolate individual cells or cell compartments from specific locations within solid tissue samples (Emmert-Buck *et al.* 1996). In combination with immunohistochemical staining, LCM allows access to cells *in situ*, which offers the opportunity to restrict expression analysis to spatially defined subareas within the tissue. Pavličev *et al.* used LCM method to enrich SCTs from fresh-frozen sectioned human placentas and analyzed SCT-specific genes (Pavličev *et al.* 2017). The limitations of LCM include insufficient extraction of mRNAs due to incomplete capture of materials, and sometimes target cells are contaminated with undesired surrounding cells. In FACS method, the targeted cells tagged with fluorophore-conjugated monoclonal antibodies can be separated from one another depending on which fluorophore they have been stained with (Julius *et al.* 1972). While FACS is capable of isolating specific cell populations with high purity and throughput, it also requires a large starting number of the cells (more than 10,000 cells) in suspension and monoclonal antibodies specific to the cells of interest. In addition, this method can introduce oxidative stress on cells, which may result in their transcriptomic changes. In MACS method, fluorophores used in FACS are replaced with magnetic nanoparticles, allowing separation of target cells when exposed to a magnetic field. MACS method is much gentler and results in better cell viability than FACS. This is particularly important to placental scRNA-seq studies because placental trophoblast cells are highly vulnerable to damage during dissociation (Liu *et al.* 2018). MACS method has been used in isolating multiple cell types from first- and second-trimester placentas (Liu *et al.* 2018). Various microfluidic techniques are used for single-cell isolation. A widely used microfluidic technique is droplet-based microfluidics, such as Drop-seq and Chromium system from 10× Genomics. In the droplet-based systems, cells are individually encapsulated into a water-in-oil droplet, followed by the cDNA synthesis via reverse transcription, during which the cell barcode and the unique molecular identifiers (UMI) are incorporated into the cDNA molecule. The cDNAs from individual cells are amplified for library preparation and subsequently next-generation sequencing (Klein *et al.* 2015; Macosko *et al.* 2015; Zheng *et al.* 2017). Droplet-based microfluidic systems are increasingly popular due to their scalability, low reagent consumption, and precise fluid control. Suryawanshi *et al.* compared two droplet-based methods, Drop-seq and 10x Chromium platforms, for scRNA-seq analysis of human first-trimester placental cells. They found that although more genes were detected in the 10x Chromium method, there was a good correlation between the Drop-seq and 10x Chromium expression data (Suryawanshi *et al*. 2018). One limitation of droplet-based methods, however, is that it requires specific microfluidic devices and instruments for droplet sorting. The commercial droplet-based platforms have limited room for optimization or modification of cell capture conditions for specific samples.

The scRNA-seq library preparation procedure involves several steps including the conversion of cellular RNA into first-strand cDNA by reverse transcription, second-strand cDNA synthesis by poly(A) tailing or by a template-switching approach, and cDNA amplification using PCR or by in vitro transcription (Chen *et al.* 2019). Then, amplified and tagged cDNAs from individual cells are pooled and sequenced. The Smart-seq2 approach generates full-length cDNA, thus has the advantage in allele-specific expression detection and alternative-splicing analysis (Picelli *et al.* 2013). For example, Vento-Tormo *et al*. used Smart-seq2 method to investigate genes of the KIR family, which are polymorphic and highly homologous, in uNK cells from early pregnancy (6-14 weeks) human placentas, and successfully mapped the single-cell reads to the corresponding donor- specific reference of KIR alleles (Vento-Tormo *et al*. 2018). Other methods combined with UMI or barcodes only capture either the 5'-end or 3'-end of the transcripts (Islam *et al.* 2011; Hashimshony *et al.* 2012). The main advantage of these methods is that they are more cost-efficient for transcriptome quantification of large cell numbers. However, these approaches are not suited for the isoform analysis and they are sub-optimal for single-nucleotide polymorphisms identification.

There are a few next generation sequencing platforms. Illumina sequencers are by far the most popular, with a claimed more than 70% dominance of the market, and have produced the majority of the publicly available human sequencing data. In 2015, BGI launched the BGISEQ-500 as an alternative to existing sequencing platforms, with lower sequencing cost. Sequencing performance comparison between the Illumina platform and the BGISEQ-500 showed that the two platforms have comparable sensitivity and accuracy in bulk RNA seq, small noncoding RNA seq, whole genome DNA seq, and scRNA-seq (Natarajan *et al*. 2019). The sequencing system chosen for a scRNA-seq experiment depends largely on the biological question being asked. Considerations for a sequencing experiment include total number of reads per cell, paired vs single read and the estimated desirable number of single cells to yield in each experiment. For example, 10x genomics recommend 50,000 read pairs per cell, with a targeted population up to 10,000 cells per sample. Single-read sequencing involves sequencing from only one end of DNA fragments in a library, while paired-end sequencing involves sequencing both ends of DNA. Single-read sequencing is faster and cheaper than paired-end sequencing and is mostly used to assess the abundancy of gene expression. Paired-end sequencing improves the ability to detect structural rearrangements (Rizzetto et al. 2017). Typically, paired-end reads (2 x 75 bp or 2 x 100 bp) are required for novel transcriptome assembly and annotation. Compared to the bulk RNA-seq method, scRNA-seq protocols have significantly more technical variations. The technical replicates can be used to assess and correct batch effects, however this is expensive given the cost of scRNA-Seq. Alternatively, some protocols use a set of synthetic RNAs with various lengths and GC content as spike-in controls (Brennecke *et al*. 2013). UMIs, which tag individual transcripts within a cell and thus can theoretically enable the estimation of absolute molecular counts, have also been used to reduce amplification bias (Kivioja *et al*. 2011). However, spike-ins and UMIs are not available for every protocol. Spike-in RNAs can be used in protocols such as Smart-seq2 but are not compatible with droplet-based approaches, whereas UMIs are usually used in methods where only the 3'-ends of transcripts are sequenced, such as Drop-seq.

Although the development of the experimental technologies and analytical methods have empowered the popularity of the single-cell RNA sequencing, one should be aware of the challenges. For example, scRNA-seq requires mechanical or enzymatic dissociation of the tissue, which could result in the loss of vulnerable cell types or even alter gene expression in cells. Thus, dissociation protocols must be optimized for the specific type of tissues of interest to minimize dissociation bias. A quality assessment of single cell preparation should be performed to ensure that cells are viable and fully dissociated into single cells before loading them onto the microfluidic chip. Two human placental scRNA-seq studies have shown that tissue dissociation may alter the frequencies of certain placental cell types (Liu *et al*. 2018; Suryawanshi *et al*. 2018). Variability in results may arise from both experimental and data analytical phases.

**Single Cell RNA-seq analysis pipelines**

The development of novel computational tools and analytical frameworks are closely coupled with single cell genomics technologies, in order to extract biological information. It is thus worthy to overview the computational pipelines and workflows, in the context of placenta scRNA-seq analysis.

There is a growing list of new computational tools and pipelines, which were reviewed in detail elsewhere (Luecken *et al*. 2019; Vieth *et al*. 2019). A typical workflow includes two stages: (1) Data preprocessing and (2) Data analysis (Poirion *et al.* 2016). In data preprocessing, raw data are processed to obtain matrices of molecular counts or read counts, depending on whether unique molecular identifiers (UMIs) were included in the library preparation protocol. This step consists of quality control, demultiplexing, alignments and quantification. Many benchmark studies have compared different aspects of the data preprocessing step and the effect of the choice of tools to downstream analysis (Vieth *et al.* 2019; Du *et al.* 2020). For example, using the splice-aware aligner STAR in conjunction with GENCODE annotation is indicated as the best practice for alignments in many studies. The data analysis step varies by the nature of the dataset and the specific goal of the study. In general, most computational pipelines include methods regarding (1) batch-correction, imputation, normalization and quality control; (2) feature selection and dimensionality reduction; (3) unsupervised clustering and supervised cell-type annotation; (4) differential expression analysis, gene set analysis and gene regulatory or protein-protein interaction network inference; (5) pseudotiming modeling, or trajectory analysis; (6) cell-cell communication analysis.

Several single-cell analysis platforms encompassing all or most of methods mentioned above are available in R (Ji *et al.* 2016; McCarthy *et al.* 2017; Butler *et al.* 2018; Street *et al.* 2018; Cao *et al.* 2019; Iacono *et al.* 2019), Python (Wolf *et al.* 2018) or as Web servers (Gardeux *et al.* 2017; Zhu *et al.* 2017) (Table 2). Platforms in programmable environments provide more flexible manipulation of the data and methods, however it requires skills in programming language and/or UNIX commands. Graphical user interface platforms are convenient for non-expert users to build a single cell analysis workflow. It has advantages for quick exploratory analysis, although some may not scale toward larger datasets.

It is worth noticing that single analytical results of cell studies often vary, due to differences in analytical framework. The increased number of bioinformatic methods proposed for analyzing and interpreting scRNA-seq data will only make unified analytical framework more challenging, if ever possible. To address the variations due to analytical methods, one approach is to enforce the same standard pipeline and re-analyze, re-annotate the samples, which will be time consuming. Alternatively, methods of data integration could be used to analyze datasets from multiple experiments on the same tissue type. A series of single-cell data integration methods have been proposed to deal with such problems, such as LIGER, Harmony, scAlign (Welch *et al*. 2019; Korsunsky *et al*. 2019; Johansen and Quon 2019). Through data integration, we could enable an increase in sample size and standardize analytical frameworks across different datasets/studies which help us generalize observations (Lähnemann *et al*. 2020). In the context of placenta study, this will enable the meta atlas for placenta tissues, benefiting the subsequent experiments especially in cell type classification and rare cell identification.

**Human placental single cell RNA-seq studies**

During pregnancy, the placenta gene expression continuously changes in order to regulate the fetal growth, maintain immunological tolerance and to modulate metabolism according to pregnancy requirement. scRNA-seq has been used to reveal heterogeneity among various cell types in the human placenta, and to elucidate molecular interactions within a population of cells in different states. Thus far, seven scRNA-seq studies have been reported on placenta, with the samples ranging from as early as $6^{th}$ week gestation to full term, in normal, preeclampsia or preterm birth conditions (Table 1).

In the first-trimester, human placenta and decidua undergo dynamic changes to enable the establishment of a successful pregnancy. Four scRNA-seq studies so far have investigated the gene expression in human placentas from early pregnancy (Suryawanshi *et al*. 2018; Liu *et al.* 2018; Vento-Tormo *et al.* 2018; Sun *et al.* 2019). Gene profiling analyses of first trimester placenta revealed multiple genes involved in cell proliferation, adhesion, differentiation, inflammatory

4responses, and metabolism, consistent with the findings from microarray study (Sitras *et al*. 2012). In addition, scRNA-seq gene profiling mapped the transcripts to the individual cells. These scRNA-seq studies also identified new cell subtypes and predicted interactions between cell types through the analysis of cell-to-cell gene expression variability.

Suryawanshi *et al*. analyzed the gene expression in more than 20,000 cells from human placental villous and decidual tissues from early pregnancy (6-11 weeks). In order to differentiate fetal and maternal origin, they only included male fetus samples. 20 distinct cell populations (9 from villi and 11 from decidua) and uncovered previously unknown subtypes of the fibroblast-like (FB) cells were identified using unsupervised clustering and differential gene expression analyses. Gene profiling showed that all cell types identified in the villi sample expressed Y chromosome–encoded genes, indicating fetal origin of these villous cells, whereas female-specific gene *XIST* was only expressed in cell types identified in the decidua samples, indicative of their maternal origin. SCTs expressed genes involved in the production of gestational hormones, including *CGA, CYP19A, HSD3B1*, and *GH2*, that support the fetal and placental development. EVTs specifically express ECM genes, consistent with their role in invasion and mobility. Three villous FB subtypes were identified, which specifically express the genes associated with endothelial cell migration, angiogenesis, and the regulation of fluid homeostasis. Two decidual FB subtypes were also identified, which expressed the genes involved in cell adhesion and lipid metabolism. These observations indicate the influence of the microenvironment on FB cell fate determination. In addition, a proliferating and a resting NK subpopulation were differentiated through gene expression and gene ontology enrichment analyses. Receptor-ligand interaction analysis between and within villi and decidua showed abundantly expressed receptors and ligands for each cell type of villi and decidua, providing insights into the cell-cell interactions in the fetomaternal microenvironment.

Vento-Tormo *et al*. investigated the transcriptome of more than 70,000 cells from the maternal–fetal interface between 6–14 weeks of gestation, and matched maternal peripheral blood cells from early pregnancy (6-14 weeks). Alignment of scRNA-seq reads with overlapping single nucleotide polymorphisms called from fetal and maternal genomic DNA showed that placental samples contained mostly fetal cells with some maternal macrophages, whereas most cells from decidual samples except a few EVTs were of maternal origin. Unsupervised clustering and differential gene expression analyses identified two clusters of decidual fibroblast cells (PV1 and PV2), three clusters of decidual stromal cells (dS1, dS2 and dS3), and three subsets of decidual NK cells (dNK1, dNK2 and dNK3). The localization of decidual fibroblast cells and stromal cells was verified by immunohistochemistry as well as multiplexed single-molecule fluorescent in situ hybridization, and the presence of decidual NK cells was confirmed by flow cytometry. PV1 cells expressed high levels of genes involved in cell adhesion, while genes associated in tissue remodeling are higher in PV2 cells. All subsets of decidual NK cells express genes encoding proteins involved in anti-inflammatory roles, which suggests a entinal role of these cells in regulating the extent of the trophoblast invasion. Expression analysis of the KIR gene family using the KIRid method indicates a likely function of dNK1 in the recognition and response to EVT. Trajectory modeling and psudotemporal ordering of cells predicted two distinct trophoblast differentiation pathways, with the decidual EVTs being at the end of the trajectory and expressing high levels of *HLA-G*. In addition, they developed a statistical tool named CellPhone to identify the specific interactions between decidual NK cells and invading fetal EVTs, maternal immune and stromal cells. Using this tool, they predicted the ligand- receptor interactions that may control the differentiation processes of trophoblast cells to SCTs or EVTs (Vento-Tormo *et al.* 2018). Placental cellular lineage commitment occurs mainly during the first and second trimester pregnancies. The process of invasion of EVTs from anchoring villi into the uterine spiral arteries is also completed by 20 to 22 weeks of gestation in normal pregnancy (Pijnenborg *et al.* 1983). Incomplete EVT invasion and failure of spiral artery transformation by this time has been reported in preeclampsia (Brosens *et al.* 1972), indicating that preeclampsia may be related to shallow EVT invasion. Liu *et al.* investigated the transcriptomic profile of sorted single cells from both first- (8 weeks of gestation) and second-trimester (24 weeks of gestation) placentas. Unsupervised clustering analysis and differential gene expression analysis of maker genes revealed 14 subtypes of placental cells which vary with the different gestational stages. Specifically, they identified three CTB subtypes (CTB_8W_1, CTB_8W_2 and CTB_8W_3) from the first trimester placenta. These CTB subtypes have various degrees of proliferative capacity, which may serve as the pool of CTBs and progenitor cells of the STBs, respectively. CTB_8W_1 cells, with the least proliferative capacity, expressed high levels of potentially important fusion genes, indicating that they may be the fusion-competent CTB cells.

They also identified five EVT subtypes (three EVT subtypes from 8-week placentas, EVT_8W, and two EVT subtypes from 24-week placentas, EVT_24W). Gene ontology analysis and immunostaining revealed that EVT_8W cells with proliferative potential were localized at the proximal end of the placental cell column, while EVT_8W cells involved in receptor activity regulation and the immune response were found to be at distal end of the placental cell column. Gene ontology analysis of EVT_24W cells showed that one subtype may be associated with the response to wounding, digestion and immunomodulation, while the other subtype related to growth regulation, gonadotropin secretion and pregnancy. A pseudotime analysis of CTB_8W, EVT_8W and EVT_24W cells predicted a differentiation pathway of CTB_8W_2 cells to CTB_8W_3 cells and then to EVT_8W and EVT_24W cells, which is consistent with the study by Vento-Tormo *et al.* (Vento-Tormo *et al.* 2018). In addition, they identified two subtypes of macrophages (Macro_1 and Macro_2) and mesenchymal stromal cells (Mes_1 and Mes_2) from the first trimester villous stromal core. Gene ontology analysis indicated a likely role of Mes_1 cells in the regulation of cell adhesion and migration, and of Mes_2 cells in the development of the mesenchyme and blood vessels. Gene ontology analysis of macrophages suggested that they may play a role in removal of dead cells or cellular debris during early placental development. Further, many polypeptide hormone genes were found to be expressed by the first-trimester SCTs, consistent with previous scRNA-seq studies (Tsang *et al.* 2017; Suryawanshi *et al.* 2018). Surprisingly, both the first- and second-trimester EVTs also expressed many hormone genes, indicating a role for these hormones in maintaining the fetal growth and the maternal adaptations to normal pregnancy (Liu *et al.* 2018). Recently, Sun *et al.* analyzed the gene expression in more than 7,000 cells from human placental villous and decidual tissues at 6-11 weeks of gestation. Using unsupervised clustering and differential gene expression analysis, they identified 5 major cell types (trophoblasts, stromal cells, hofbauer cells, antigen-presenting cells and endothelial cells) with unique crosstalk at the fetal-maternal interface. Gene ontology enrichment analysis showed that the top enriched processes include cell proliferation and adhesion in trophoblasts, cell differentiation and signal transduction in stromal fibroblasts, inflammatory responses in Hofbauer cells, and cell migration and angiogenesis in endothelial cells. Cell-type specific upstream regulators were predicted by Ingenuity pathway analysis. In addition, a number of chemokine genes were found to have sex-differential expression in trophoblasts and Hofbauer cells, suggesting a sexual dimorphism of immune responses in the placenta (Sun *et al*. 2019). Although both microarray and bulk RNA-seq studies have shown sex-biased gene expression in human placentas at first-trimester and term (Sood, *et al*. 2006; Buckberry *et al*. 2014; Gonzalez *et al*. 2018), this is the first report by far that has investigated the sex-specific differences in single cell transcriptome of the placental cells. Given the vast differences in the types of cells and cell subtypes identified in these studies, it will be of interest to perform meta-analysis and comparative studies, to confirm the cell types that are either claimed novel, as well as the degrees of consistency among common cell types.

Three studies have examined the human placental transcriptome during the third trimester. Pavličev *et al*. analyzed the gene expression profiles of full-term placental cells. They compared the single-cell transcriptomes with the tissue-level data and showed that scRNA-seq analysis recovered more than 4000 genes that were not detected in the tissue-level transcriptome, consistent with the view that scRNA-seq may reveal the expression profiles of rare cells. They identified five distinct clusters belonging to different types of decidual and fetal cells of utero-placental interface using hierarchical clustering and marker gene expression analysis. Among the five trophoblast clusters, the large clusters were identified as intravillous cytotrophoblast (iVCT), and the smaller cluster as EVT. These trophoblast clusters likely reflect the continuous differentiation process of cytotrophoblasts. The expression profiles of iVCT cells suggest that they are a very heterogeneous group and that the differentiation proceeds by individual cells in a different order. In addition to the identification of cell clusters, they inferred the cell-cell communication network at the maternal-fetal interface using public databases of ligand-receptor relationships based on the mRNA expression. A large number of interactions were found to be between decidual cells and the SCT cells. Further, they showed that the cell-type specificity of G-protein-coupled receptor (GPCR) profiles may serve to biological characterize the cell type identity.

Tsang *et al*. analyzed single-cell transcriptomes of more than 24,000 cells isolated from full-term and second trimester preeclamptic placentas. Clustering analysis by t-SNE identified diverse placental cell types and comparison of the ratio of transcripts carrying fetal-specific SNP alleles with those carrying maternal-specific SNP alleles showed that most clusters were of fetal origin. Gene profiling analysis showed that SCTs specifically expressed genes involved in the production of important gestational hormones, and EVTs expressed HLAs associated with maternal immune tolerance of the fetus with uNKs. In addition, they predicted the differentiation relationship of trophoblast cells using reclustering and pseudotime analyses. They showed that SCTs expressed high levels of the genes involved in cell fusion and identified several genes that could serve as regulators in SCT development. Analysis of the EVT developmental pathway revealed up-regulation of various extracellular enzymes associated with invasion and migration. Importantly, comparison of transcriptomic profiling of biopsies taken from different sites of the placenta showed spatial heterogeneity, consistent with previous bulk tissue transcriptomic profiling (Sood *et al*. 2006). Furthermore, they integrated the single-cell-derived placental signature genes with maternal plasma-circulating RNA analysis to analyze the fetal and maternal cellular dynamics during pregnancy progression and suggested that the cellular pathology in early preeclampsia can be identified noninvasively (Tsang *et al.* 2017).

More recently, Pique-Regi *et al*. carried out a comparative study to define individual cell-type–specific gene signatures during physiologic and preterm labor (Pique-Regi *et al.* 2019). They analyzed cells from placental villi, basal plate, and chorioamniotic membranes of women with or without labor at term (38-40 weeks of gestation) and those with preterm labor (33-35 weeks of gestation). Using differential gene expression analysis of marker genes, a new cluster of non-proliferative interstitial cytotrophoblasts was identified in the placental villi, and the basal plate which is adjacent to the placental villi. A new cluster of lymphatic endothelial decidual cells was also revealed in the chorioamniotic membranes using cell-type–specific gene signatures. Gene enrichment analysis using *clusterProfiler* indicated that these lymphatic endothelial decidual cells may mediate the influx of immune cells into the chorioamniotic membranes. In addition, significant differences in transcriptional profiles were identified among placental cell types and between the term and preterm labor, with the largest number of differentially expressed genes in the maternal macrophages between the term labor and term no labor groups. Further, an in silico analysis in public datasets showed that the single-cell signatures of macrophages, monocytes, activated T cells, and fibroblasts is modulated in the circulation of women with preterm labor compared to gestational age-matched controls. These results suggested the feasibility of using single-cell signatures as biomarkers to non-invasively monitor the complex fetal and maternal cellular dynamics for normal or pathological conditions in pregnancy. Yet, additional studies are needed to characterize the transcriptional profiles of the very preterm labor (less than 32 weeks of gestation).

**Perspectives and conclusions**

Placenta is a very heterogeneous, blood-vessel rich organ with maternal and fetal origins and various cell types. The human placenta scRNA-seq studies reviewed here reveal that vast transcriptional differences exist among geolocations of placentas, between various cell types, gestational ages as well as in pregnancy diseases. These studies open up the possibilities to better understand cellular networks and pathways in cells of normal and complicated pregnancies, which could help to discover the underlying causes of adverse pregnancy outcomes.

However, one needs to be cautious before attempting scRNA-Seq studies using human placenta. Access to normal human placenta before delivery is challenging due to ethical concerns, limiting the kind of studies that can be conducted. Upon delivery, various factors need to be taken in consideration including hospital procedures, mode of delivery and sample sizes. Placenta grows and constantly adapts to the maternal environment to ensure a healthy pregnancy, thus static views of placenta cell proportions and gene expression are incorrect. Temporal, spatial and between individual variability have to be considered. Fresh samples are limited to deliveries and planning is critical. This also comes with challenges regarding deconvoluting disease state vs. batch variations. To get around these issues, single-nucleus RNA sequencing (snRNA-seq) has arisen as an alternative to single-cell methods. snRNA-seq does not require enzymatic digestion, which reduces dissociation artifacts. In addition, unlike scRNA-seq which requires the preparation of a live single cell suspension from fresh tissue which inevitably introduces batch effect, snRNA-seq can be used for frozen tissues, enabling single-cell gene profiling of archived clinical samples simultaneously, reducing batch effect and increasing power of the study by allowing for more biological replicates (Habib *et al*. 2017).

Despite these technological and practical challenges, in the coming years, we expect to see more studies that will reveal the normal placenta developments throughout normal pregnancy and provide a more complete picture for a dynamic placenta atlas, such as those in Pediatric Cell Atlas (PCA) (Taylor *et al*. 2019). More single cell studies to investigate the molecular details of abnormalities in other pregnancy complications, such as gestational diabetes, could be interesting.

23scRNA-seq studies that address the effect of lifestyle choices, such as smoking and alcohol use in pregnancy, will also pingpoint the biological changes in placentas at single cell level.

Additionally, with the development of single cell technologies, other types of single cell omics studies, such as single cell epigenetics studies, as well as multi-omics single cell studies (scRNA-seq, sc methylation and scATAC-Seq etc) will be very useful to delineate the coherent changes from epigenetic to transcriptomics level (Poirion *et al.*; Ortega *et al.* 2017). For example, single-cell triple omics sequencing (scTrio-seq) (Hou et al., 2016), is a powerful technology that can simultaneously identify copy-number variations (CNVs), DNA methylome, and transcriptome of the same single mammalian cell. It was used to sequence 25 single cancer cells derived from a human hepatocellular carcinoma tissue sample and identified 2 subpopulations. The same team later applied this technology to colorectal cancer tumors and metastases from 10 individual patients, and showed that DNA methylation levels are consistent within lineages but differ substantially among clones (Bian et al., 2018). For more detailed experimental methods, the readers can refer to comment recently (Zhu et al., 2020). Another new technology that can potentially help understand the heterogeneity of placenta *in situ*, is the spatial transcriptomics, where single cell (or close to single cell ) resolution of gene expression data can be spatially mapped back to the tissue at 2D level (Moncada *et al.* 2020), where the 2D images can be stacked to reconstruct 3D images of the tissue.

From the data analytics perspective, given the unique challenges of placenta, computational methods that can help automatically excluding blood cells due to experiments, decipher the maternal vs. fetal cell origins, and automatically annotate cell types in placenta will be critical for understanding the physiology and pathology of placenta. Besides pipelines and workflows for single cell RNA-Seq, tools and methods that aim at meta-analysis or comparative analysis of different datasets will be valuable to uncover the development trajectory of placenta through gestational ages. Methods that integrate single cell multi-omics and multi-modal data, such as those handling spatial transcriptomics data, will be much needed for deeper characterization of placenta tissue. It will be interesting to see if and how the multi-omics data integration methods developed for bulk cell analysis can be extended to single cell (Huang *et al.* 2017). As the field moves forward with exponential growth of single cell genomics data, all methods will need to strive for accuracy and scalability, in order to be truly useful and withstand the test of time (Garmire *et al.* 2018). In the end, advances in single-cell transcriptomics combined with computational analysis tools will enable significant improvement in both basic and clinical research related to the human placenta.

**Author contributions**

LXG envisioned and supervised the project. HL wrote the majority of the manuscript, with significant help from QH, YL and LXG.

**Declaration of interest**

The authors declare that there is no conflict of interest that could be perceived as prejudicing the impartiality of this review.

**Funding**

This work was supported by grants K01ES025434 awarded by NIEHS through funds provided by the trans-NIH Big Data to Knowledge (BD2K) initiative (www.bd2k.nih.gov), R01 LM012373 and R01 LM012907 awarded by NLM, and R01 HD084633 awarded by NICHD to L.X. Garmire.

**Figure Legend**

Figure 1. Single cell RNA-sequencing on placenta cells. The heterogeneity of placenta gene expression can be revealed by single cell RNA-sequencing (scRNA-seq) technology. The placenta tissue is first dissociated into a single cell suspension and cells are then individually encapsulated into a water-in-oil droplet, followed by the cDNA synthesis via reverse transcription, during which the cell barcode and the unique molecular identifiers (UMI) are incorporated into the cDNA molecule. The cDNAs from individual cells are amplified for library preparation, followed by next-generation sequencing. The sequences are then mapped by alignment algorithms and counted for those mapped to the reference transcriptome. The cell origin of the mapped reads can be identified by the cell barcode within them. As a result, a sequence read count matrix is generated over thousands of genes (rows) and thousands of single cells (columns). This count matrix is then subject to preprocessing and downstream bioinformatics analysis, such as clustering and pseudo-time reconstruction among single cells, as shown in the figure. EVT, extravillous trophoblast; F, fibroblast; H, Hofbauer cell; SCT, syncytiotrophoblast; VCT, villous cytotrophoblast.

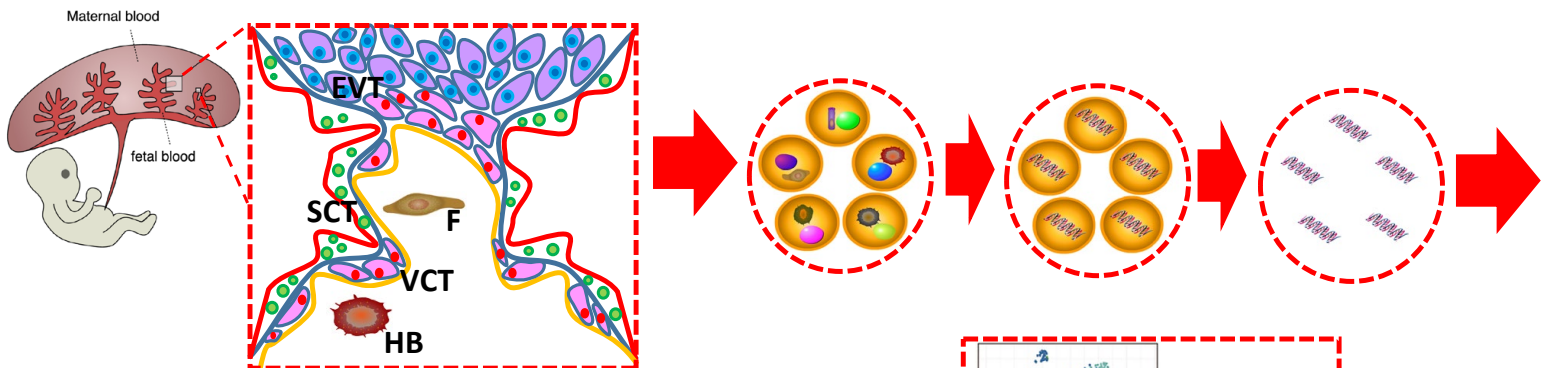
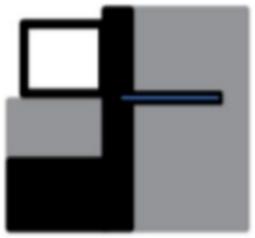
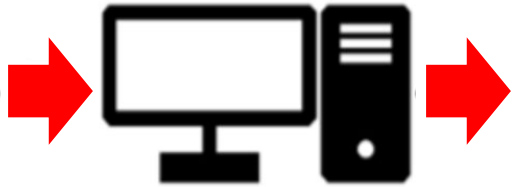
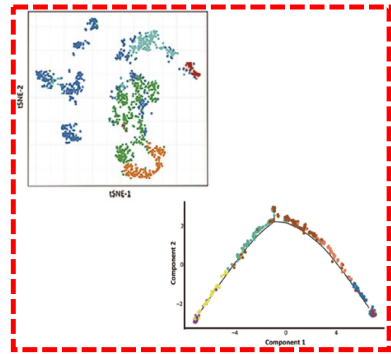

**Table 1. Summary of current single cell RNA-seq studies on human placenta**

| Gestational age (weeks) | Tissue | Number of cells | Number of placenta | Single cell isolation | Platform | Average reads length (bp) | Mean Pair-end reads | Reference |
|---|---|---|---|---|---|---|---|---|
| 6-11 | Villous, decidual | 14,341 6,754 | 9 | Enzyme | 10x Drop-seq | 100 | N/A | Suryawanshi *et al*. (2018) |
| 6-14 | Placenta, blood | 70,000 | 5 | Enzyme, FACS | 10x Smart-seq2 | 75 | $1\times10^6$ | Vento-Tormo *et al*. (2018) |
| 8, 24 | Placenta | 1,567 | 8 | Enzyme, gradient centrifugation, MACS | Smart-seq2 | 150 | $1\times10^6$ | Liu *et al*. (2018) |
| 11-13 | Placenta | 7245 | 10 | Enzyme | 10x | N/A | N/A | Sun *et al*. (2019) |
| Term | Placenta | 87 | 2 | Enzyme, , gradient centrifugation | Fluidigm | 75 | $3.65\times10^6$ | Pavličev *et al*. (2017) |
| Term (preeclampsia or healthy | Placenta, blood | 24,000 | 8 | Enzyme | 10x | 130 | 35800 | Tsang *et al*. (2017) |
| Term (with or without labor) preterm (33-35 weeks) | Placenta | 79,906 | 9 | Enzyme | 10x | N/A | N/A | Pique-Regi *et al*. (2019) |

**Table 2. A list of platforms for single cell RNA-Seq analysis**.

| Tool | Platform | URL | Description | Reference |
|---|---|---|---|---|
| Granatum | Web | http://granatum.dcmb.med.umich.edu:8102/?_state_id_=6e60464c6d7d66a3&tab=info | A web-based comprehensive scRNA analysis framework with easy utilization | Zhu *et al*. (2017) |
| GranatumX | Web | http://35.222.3.147:34567/step/f6ba5588-93c1-4507-8013-196273fb055b | Updated version of Granatum with customized analysis modules | Zhu *et al*. (2018) |
| ASAP | Web | https://asap.epfl.ch/ | Web-based automatic single-cell analysis pipeline | Gardeux *et al*. (2017) |
| Seurat | R | https://satijalab.org/seurat/ | A comprehensive scRNA-seq analysis framework, includes the anchor technology for data integration | Butler *et al*. (2018) |
| Monocle3 | R | https://cole-trapnell-lab.github.io/monocle3/ | A comprehensive scRNA-seq analysis framework, has good performance on pseudotime cell trajectories construction | Cao *et al*. (2019) |
| bigSCale2 | R | https://github.com/iaconogi/bigSCale2 | A comprehensive scRNA-seq analysis framework, especially appropriate for large datasets | Iacono *et al*. (2019) |
| Scater | R | https://bioconductor.org/packages/release/bioc/html/scater.html | A comprehensive scRNA-seq analysis framework, with a focus on quality control and visualization | McCarthy *et al*. (2017) |
| Slingshot | R | https://bioconductor.org/packages/release/bioc/html/slingshot.html | Designed to model developmental trajectories in single-cell RNA sequencing data | Street *et al*. (2018) |
| TSCAN | R | https://github.com/zji90/TSCAN | Designed to model developmental trajectories in single-cell RNA sequencing data | Ji *et al*. (2016) |
| Scanpy | Python | https://icb-scanpy.readthedocs-hosted.com/en/stable/ | A scalable scRNA-seq analysis toolkit, able to efficiently deal with datasets of more than one million cells, because of the Python-based implementation | Wolf *et al*. (2018) |